# Photon-Ring Retarded-Time Tomography: The Mahakal Phenomenon

Historical Reach, Identifiability, and Bounded Recovery from Higher-Order Images

Hina Dixit Abhinav Chauhan

ABSTRACT

Near-critical null geodesics generate image-order channels with distinct mappings, delays, and weights, so one observer-time data set samples multiple source epochs. We organize the inverse problem by historical reach, dimension, and held-out recovery. Using continuum-noise-consistent, matrix-free Kerr/Schwarzschild operators for orders $n = 0, 1, 2$, we audit 12 spin–inclination geometries. At normalized $\mathrm{SNR}_0 = 100$, resolved orders extend anchor-connected history beyond direct imaging in all cells by $40M$, $24M$, and one $4M$ grid step at inclinations $20^\circ$, $50^\circ$, and $75^\circ$. Operational rank rises from 153 to 202 and old-age innovation from $8.29M$ to $40.72M$; complete order summation retains only $9.66M$.

Compact temporal support gives an exact result: temporal basis functions outside the direct retarded-time footprint generate identically zero direct columns, while higher orders can lift those old-epoch blind blocks. Direct old-structure rank is zero, while resolved orders support 38, 93, and 185 operational directions across three classes. Yet source enrichment leaves reach nearly fixed while rank fraction and singular-value stability collapse—the Mahakal phenomenon.

Held-out tests show bounded benefit, not a movie. Resolved orders extend stable baseline-inclusive emissivity-level span from $48M$ to $80M$, but that endpoint is level dominated. On 60 histories, physical morphology error falls by 13.3% with TSVD and 16.4% with ridge. Reliable two-feature recovery fails and the stable morphology interval is zero. The positive results are therefore best-case, known-geometry image-domain benchmarks under ideal order separation; $\mathrm{SNR}_0$ is an operator normalization, not an observational forecast.

## 1 Introduction

Light emitted near a black hole can reach a distant observer directly or after one or more near-critical windings around the photon shell. Higher-order images are demagnified and approach the critical curve, but they also carry larger, distributed delays [1, 2]. For a variable source, the same observer-time data set therefore combines source epochs. Slow-light calculations, glimmer, echoes, and autocorrelation studies establish that this time-domain structure is physically present [3–7]. The inverse question is harder: which parts of the earlier source are actually determined by the data?

We organize that question around three headline quantities:

1. **Historical reach:** how far back does the likelihood contain detectable information about a localized source perturbation?
2. **Historical dimension:** how many independent old-source directions remain determined at the declared SNR and source resolution?
3. **Historical recovery:** does a held-out estimator reconstruct the declared historical object under a prespecified error criterion?

Causal arrival is a prerequisite to all three, but it is not itself recovery. Likewise, exact rank, finite-SNR observability, and estimator performance are distinct stages of the argument.

> DEFINITION 1.1 — THE MAHAKAL PHENOMENON
>
> In a retarded-time black-hole inverse problem, higher-order geodesic footprints can enter source-time regions that the direct footprint does not reach, thereby creating access to earlier history. Yet the temporal localization and source-model enrichment required to represent that history also expose exact or finite-SNR blind directions. We call this geometry-specific creation–destruction relation the Mahakal phenomenon: historical reach can increase while recoverable dimension, identifiability, and reconstruction stability decrease. The name does not denote a new law of null-geodesic propagation or a metaphysical property of black holes.

The central question is therefore:

> Which finite, explicitly declared components of an exterior source history are encoded, stable at finite SNR, and recoverable from direct and higher-order black-hole images?

TABLE 1

Working terminology and claim boundaries. Reach, dimension, and recovery are the three organizing questions; the remaining rows are diagnostics or target definitions used to answer them.

| Term | Symbol or example | What it measures | What it does not establish |
|---|---|---|---|
| Historical reach | $L_{\mathrm{anchor}}, T_{\mathrm{reach}}$ | How far into retarded age the data contain threshold-level sensitivity. | How many directions are supported, or whether an estimator recovers them. |
| Anchor-connected span | $L_{\mathrm{anchor}}$ | The contiguous passing interval beginning at the youngest fully supported age. | The deepest isolated passing age. |
| Localized directional reach | $T_{\mathrm{reach}}$ | The deepest age at which at least one normalized localized direction passes. | A contiguous historical interval. |
| Historical dimension | $D_{\mathrm{hist}}$, operational rank | The number of independent historical directions supported at the declared SNR. | Continuum injectivity or reconstruction accuracy. |
| Old-age innovation | $J_{\mathrm{old}}$ | Integrated finite-SNR sensitivity beyond the direct-order age boundary. | Correct physical pairing of delay, position, and weight. |
| Historical recovery | held-out error, $L_{\mathrm{stable}}$ | Whether a fixed estimator recovers the declared source object on unseen histories. | Recovery outside the tested representation or source families. |
| Class-conditional error | error to best in-class projection | Estimation error after removing representation mismatch. | Physical end-to-end fidelity to the analytic source. |
| Physical end-to-end error | error to analytic source | Combined representation and inversion performance. | Instrumental realism, geometry robustness, or a stable movie interval. |

## 1.1 Contributions

The paper makes four connected contributions. First, it formulates a distributed-delay, order-resolved forward operator with continuum-noise-consistent whitening and explicit order mixing. Second, it gives a compact information-boundary theory whose distinctive result is an exact direct-image old-epoch blindness theorem for temporally localized source spaces. Third, it validates the operator on a prespecified 12-cell Kerr/Schwarzschild grid and separates the roles of delay diversity, spatial remapping, order labels, attenuation, and source-class enrichment. Fourth, it reports two sealed held-out inverse results—stable emissivity-level history extension and reduced aggregate morphology error—with their negative boundaries reported beside them.

### 1.2 Scope

"Multi-geometry" means the complete prespecified 12-cell audit, not the continuous Kerr parameter space. Held-out inversion is performed only at $a_\star = 0.5, i = 50^\circ$. The observation model is ideal, monochromatic, scalar-intensity, and image-domain. The positive morphology result requires separately resolved image orders; complete order summation does not reproduce it. No result is claimed for current Event Horizon Telescope data, unknown geometry, sparse Fourier sampling, arbitrary source movies, or continuum injectivity.

## 2 Relation to Existing Work

The near-critical Kerr image hierarchy and its interferometric signatures are established [1, 2]. Time-domain studies use glimmer, echoes, autocorrelations, or image correlations to infer lensing structure and black-hole parameters [3–6, 8–11]. Those works demonstrate repeated and delayed image structure. Here the likelihood is treated as an inverse operator, and the target is the recoverable source history conditional on a declared source space.

AART supplies the adaptive analytic Kerr ray tracing used for the primary maps [12]; kgeo provides an independent cross-tracer. Slow-, brisk-, and fast-light comparisons quantify how delay distributions interact with intrinsic source timescales [7]. Photon-ring overlap has also been analyzed directly for static spherically symmetric spacetimes and thin disks [13]; that work reinforces that ideal image-order separation is a nontrivial observational assumption, although its geometry and source model do not define the leakage operator studied here.

Dynamic black-hole imaging and tomography infer constrained source or geometry parameters from time-variable measurements [14–16]. Modern video inverse methods can impose powerful spatiotemporal priors [17, 18]. Such priors may stabilize or select a movie after a forward model is specified; they do not establish which historical components are supported by the likelihood. Our emphasis is therefore the information boundary: retarded-time footprints, null spaces, finite-SNR spectra, source-class stress, and held-out controls.

The mathematical framework is related to Lorentzian light-ray transforms and regularization of ill-posed inverse problems [19–24]. The distinct object here is an image-order-resolved source-history operator in a fixed black-hole geometry, together with age-local and resolution-aware recovery criteria.

## 3 Retarded-Time Forward Model

### 3.1 Continuum model

Let $g$ denote a stationary Kerr or Schwarzschild metric and let

$$j = j(r, \phi, t, \nu, \mathcal{P}) \tag{1}$$

be a bounded exterior emissivity. We begin with an equatorial, optically thin, monochromatic, scalar-intensity model and suppress $\nu$ and polarization. Let $\xi = (\alpha, \beta)$ denote an observer-screen coordinate and $t_o$ an observer time. For image order $n$, define the transfer map

$$\mathcal{T}_{g,n}\, j(\xi, t_o) = \chi_{g,n}(\xi)\, w_{g,n}(\xi)\, j\big(r_{g,n}(\xi), \phi_{g,n}(\xi), t_o - \Delta_{g,n}(\xi)\big) \tag{2}$$

where $\chi$ is a validity mask, $w$ is the declared transfer coefficient, and $\Delta$ is the coordinate-time delay relative to the repository convention. For specific intensity, the ray-wise coefficient includes the redshift factor $g^3$; pixel area enters the measurement and noise model rather than being silently absorbed into a scalar order amplitude.

Throughout, $n = 0$ denotes the direct image band and $n = 1, 2$ the first two retained indirect lensing bands under the ray-map classification. These are operational image-order labels: individual rays within a band have distributed delays and winding angles, so the notation does not assert that every ray executes exactly $n$ complete orbits.

The order-resolved observation is

$$d_R = \mathcal{A}_{g,N}\, j + \eta_R, \qquad \mathcal{A}_{g,N}\, j = \begin{pmatrix} \mathcal{T}_{g,0}\, j \\ \mathcal{T}_{g,1}\, j \\ \vdots \\ \mathcal{T}_{g,N}\, j \end{pmatrix} \tag{3}$$

A linear order mixer $L$ produces

$$d_L = (L \otimes I) d_R, \qquad C_L = (L \otimes I) C_R (L^\top \otimes I) \tag{4}$$

The unresolved spatial image and total-flux readout are special cases with different linear collapse operators. Their noise must be propagated through the same transformation; assigning each arm an independent scalar noise level creates artificial information.

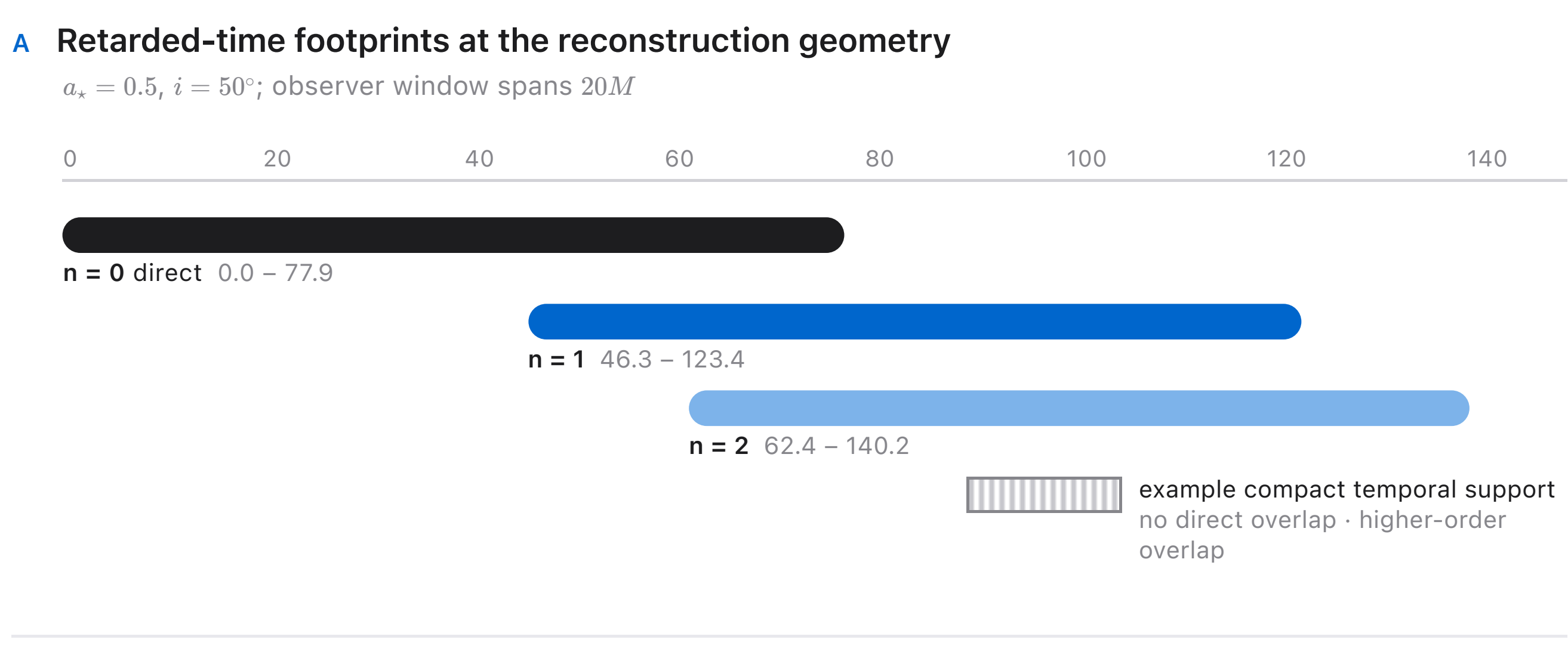

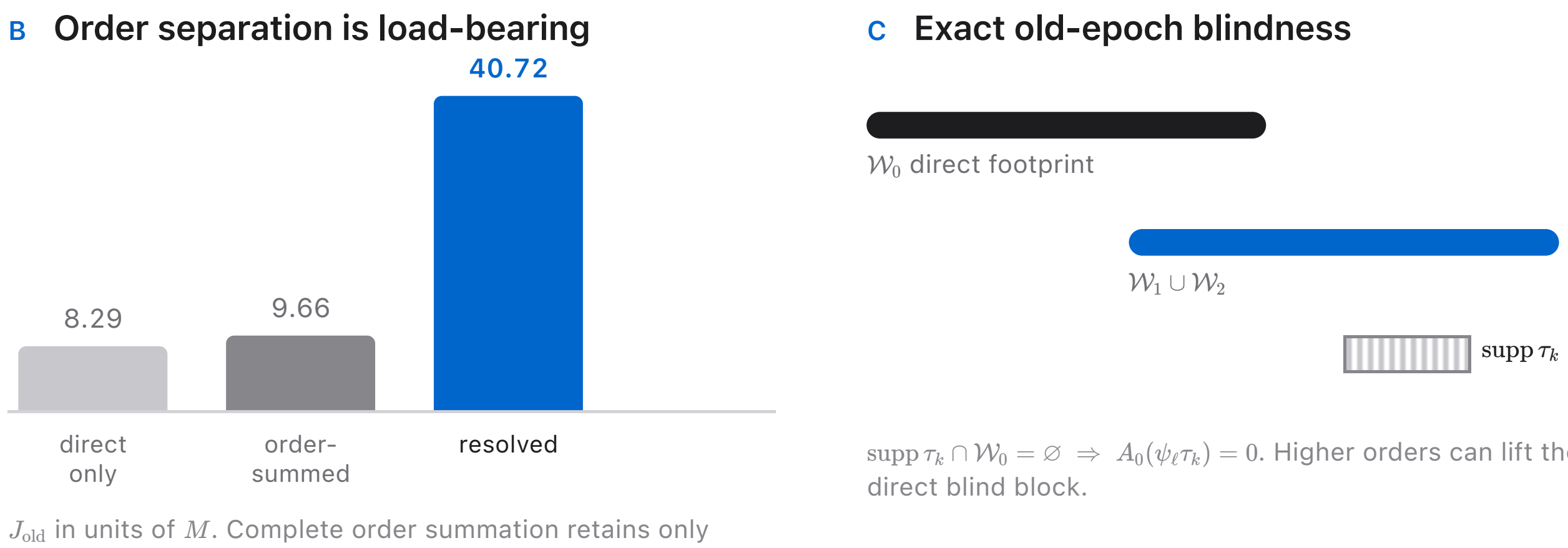

FIGURE 1

Deterministic schematic built from the reference-geometry values. (A) With an observer window spanning $20M$, the three retained order footprints cover progressively older source ages; a compact temporal mode can miss the direct footprint while intersecting higher-order footprints. (B) Complete order summation retains only $9.66/40.72 = 23.7\%$ of the resolved old-age innovation. (C) The support condition underlying the direct-image old-epoch blindness theorem in Theorem 4.1. The figure is explanatory but uses the numerical delay and information values reported below.

### 3.2 Linear representation spaces and bounded source sets

Let $Q_d : \mathbb{R}^d \to \mathcal{X}$ be a linear synthesis map and define the finite-dimensional model space

$$\mathcal{S}_d = \mathrm{range}(Q_d) \tag{5}$$

As a vector space, $\mathcal{S}_d$ is not bounded. When a deterministic secant-stability statement requires a bounded source set, we instead choose an explicitly declared compact coefficient set $K \subset \mathbb{R}^d$ and write

$$\mathcal{C}_d(K) = \{Q_d c : c \in K\} \subset \mathcal{S}_d \tag{6}$$

The continuum operator $\mathcal{A}_{g,N} : \mathcal{X} \to \mathcal{Y}$ and its restricted coefficient matrix are distinct objects:

$$A^{(d)}_{g,N} = \mathcal{A}_{g,N} Q_d \tag{7}$$

Rank, singular values, and conditioning reported below are properties of $A^{(d)}_{g,N}$ at declared tolerances; they do not establish injectivity of the continuum operator $\mathcal{A}_{g,N}$.

The physical representation is

$$\begin{gathered} \mathcal{S}_{224} = \mathrm{range}(Q_{224}), \\ \text{4 radial B-splines} \times \text{7 real azimuthal modes} \\ \times \text{ 8 temporal DCT-II modes} \end{gathered} \tag{8}$$

The radial functions are clamped cubic B-splines on uniform knots in $\log r$; the azimuthal functions are $1$, $\cos m\phi$, and $\sin m\phi$ for $m = 1, 2, 3$; and the temporal functions are eight global cosine modes on the declared source-time window. Repository configurations retain the historical identifier C224; throughout the manuscript, $\mathcal{S}_{224}$ denotes the linear representation space and $\mathcal{C}$ is reserved for explicitly bounded source sets.

For deterministic secant statements, $K$ must genuinely be compact. The phenomenological generators below include Gaussian random weights and are therefore not treated as a compact continuum class. For the sealed main experiment, let $\mathcal{B}_{\mathrm{in}}$ denote the 320 exact-in-representation held-out histories (256 fitting-family histories and 64 held-out-flare histories), with stored coefficient vectors $c_j$. We define the finite coefficient set

$$K^{\mathrm{level}}_{224} = \{c_j : j \in \mathcal{B}_{\mathrm{in}}\}, \qquad \mathcal{C}^{\mathrm{level}}_{224} = Q_{224} K^{\mathrm{level}}_{224} \subset \mathcal{S}_{224} \tag{9}$$

Because it is finite, $K^{\mathrm{level}}_{224}$ is compact. Rank and singular-value claims apply to $A^{(224)} = \mathcal{A} Q_{224}$ on the linear space $\mathcal{S}_{224}$; deterministic secant statements apply to explicitly bounded sets such as $\mathcal{C}^{\mathrm{level}}_{224}$; and the reconstruction result is empirical over the sealed bank. It is not a uniform guarantee over a continuous source-family distribution, the convex hull of the bank, or every coefficient vector in $\mathbb{R}^{224}$.

### 3.3 Age-localized physical probe

The global DCT modes measure coefficient identifiability averaged over the full time window; they do not localize a particular epoch. The physical historical probe is therefore defined directly in source-function space. For retarded age $a \geq 0$,

$$q_a(r,\phi,t) = \frac{\mathbf{1}_{[r_{\text{in}},r_{\text{out}}]}(r)}{N_h} \exp\left(-\frac{(t+a)^2}{2h^2}\right), \qquad h = 3M \tag{10}$$

with no azimuthal modulation. Under the equatorial area measure,

$$N_h^2 = \pi(r_{\text{out}}^2 - r_{\text{in}}^2)h\sqrt{\pi} \tag{11}$$

so every probe has unit $L^2$ norm over the emission annulus. For observation arm $r$,

$$\mathcal{I}_g^{(r)}(a) = \left\|C_{g,r}^{-1/2}\mathcal{A}_{g,r}q_a\right\|_2^2 \tag{12}$$

If a probe lies in a chosen representation, this equals the corresponding coefficient quadratic form. The physical probe is evaluated directly rather than projected through the global temporal DCT basis, so it has no oscillatory projection side lobes.

The physical audit evaluates centers on a $4M$ age grid and uses $\rho = 1$ in the detectability condition. For anchor bookkeeping, the operational probe support is $|t+a| \leq 3h$, which contains $1 - \text{erfc}(3) \simeq 0.999978$ of the probe's squared $L^2$ energy. The Gaussian tails remain part of the forward evaluation; the $3h$ convention only determines whether an age center is fully supported by the reachable source-time window. The full threshold mask is retained, so side gaps and endpoint effects are not compressed into one oldest-age number.

For an observer window $[t_{o,\min}, t_{o,\max}]$, order $n$ reaches the source-time footprint

$$\mathcal{W}_n = \left[t_{o,\min} - \Delta_n^{\max},\ t_{o,\max} - \Delta_n^{\min}\right] \tag{13}$$

Widening the observing window increases absolute reach even for the direct image; the order-specific old-edge excess is instead controlled by differences such as $\Delta_n^{\max} - \Delta_0^{\max}$. All reported age endpoints inherit the fixed $h = 3M$ probe and $4M$ age grid.

No independent $h$-sensitivity study was performed, so a one-bin $4M$ gain should be read as a threshold-scale extension rather than a sharply resolved historical layer.

The scalar $q_a$ is spatially flat. For the delay-versus-spatial mechanism audit we also use a 28-direction localized class

$$q_{a,\ell}(r,\phi,t) = \psi_\ell(r,\phi)\, N_{h,\ell}^{-1} \exp\left(-\frac{(t+a)^2}{2h^2}\right), \qquad \ell = 1,\ldots,28 \tag{14}$$

where $\psi_\ell$ are the four radial factors crossed with the seven real azimuthal factors. Each direction is normalized independently, and the age-local Fisher matrix is $M(a) = P(a)^\mathsf{T} P(a)$ for the whitened response matrix $P(a)$. This localized class is needed because a spatially flat scalar probe is algebraically invariant under a spatial-only substitution and cannot by itself diagnose the mechanism.

Because detectability depends on the ratio $\rho/\mathrm{SNR}_0$, multiplying $\rho$ by a factor $c$ is equivalent to dividing $\mathrm{SNR}_0$ by $c$. The reported SNR sweep therefore also supplies threshold sensitivity for this scalar criterion. At the reconstruction geometry, the equivalent threshold sweep at fixed $\mathrm{SNR}_0 = 100$ is shown in Table 2. The old-age innovation $J_{\mathrm{old}}$ is reported separately because it does not use a detection threshold.

TABLE 2

Threshold sensitivity of the scalar age endpoint at $a_\star = 0.5$, $i = 50^\circ$. The middle row is the primary $\rho = 1$ convention. The other rows are read from the same fixed SNR sweep using the equivalence $\rho/\mathrm{SNR}_0$.

| $\rho$ at $\mathrm{SNR}_0 = 100$ | Equivalent sweep SNR | Direct span ($M$) | Resolved span ($M$) |
|---|---:|---:|---:|
| 3.33 | 30 | 56 | 56 |
| **1.00** | **100** | **60** | **84** |
| 0.333 | 300 | 60 | 96 |

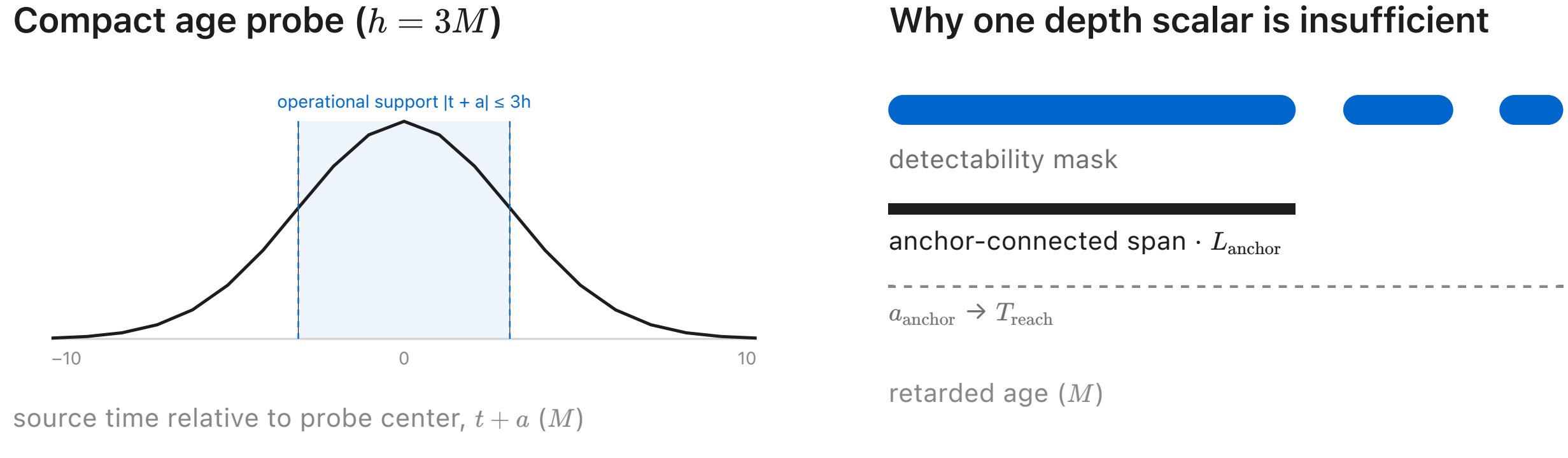

FIGURE 2

Age probe and interval semantics. Left: the unit-normalized Gaussian source-function probe used in the physical audit; the $3h$ support convention is used only for anchor bookkeeping, while the full Gaussian enters the forward evaluation. Right: a schematic disconnected detectability mask illustrating why the oldest passing age, the longest passing run, and the run connected to the physical anchor are distinct observables.

### 3.4 Continuum image-noise convention

The primary image-domain model uses white specific-intensity noise density per unit solid angle. For a pixel-average datum with solid angle $\Delta\Omega_p$,

$$\mathrm{Var}(\epsilon_p) = \frac{\sigma_\Omega^2}{\Delta\Omega_p} \tag{15}$$

The corresponding whitened row of a restricted physical matrix is

$$\tilde{A}^{(d)}_{p,:} = \frac{\sqrt{\Delta\Omega_p}}{\sigma_\Omega}\, g_p^3\, B_d(r_p, \phi_p, t_p) \tag{16}$$

Artificially splitting a pixel into equal-area children with identical local transfer values must leave the Gram matrix invariant. The relative split/merge discrepancy must remain below $10^{-10}$. This prevents finer quadrature from receiving a fictitious SNR advantage.

### 3.5 Reference-SNR normalization

The dimensionless $\mathrm{SNR}_0$ is a leading-order operator normalization, not a per-pixel, peak-flux, total-flux, or current-telescope SNR. Operators are first assembled at $\sigma_\Omega = 1$. Let $j_{\text{ref}}(r, \phi, t) = 1$ be the constant reference emissivity, which lies in $\mathcal{S}_{224}$ through the B-spline partition of unity and the constant azimuthal and temporal modes. If $m_0$ is the number of direct-order data rows, define

$$s_{\mathrm{ref}} = \left( \frac{1}{m_0} \left\| \mathcal{A}_{g,0}^{(\sigma_\Omega=1)} j_{\mathrm{ref}} \right\|_2^2 \right)^{1/2} \tag{17}$$

A sweep point $\mathrm{SNR}_0$ is obtained by using

$$\sigma_\Omega = s_{\mathrm{ref}} / \mathrm{SNR}_0 \tag{18}$$

Thus $\mathrm{SNR}_0 = 100$ means that the RMS whitened response of the direct-order operator to the constant reference source is 100. This one noise density is then held fixed for every observation arm within that geometry; unresolved and total-flux covariances are obtained by linear propagation rather than by assigning those arms a new noise level.

NORMALIZED-SNR CONVENTION

$\mathrm{SNR}_0$ is an operator-normalization parameter. It is not a per-pixel SNR, total-flux SNR, EHT image SNR, or forecast of current observing feasibility. For order $n$, the reference-source response is $S_n = \|C_n^{-1/2} A_n j_{\mathrm{ref}}\|_2$ and the RMS response per retained datum is $S_n/\sqrt{m_n}$; for a general source direction $v$, directional SNR is $\mathrm{SNR}_0 \|B_n v\|_2$.

In the 12-cell audit, $s_{\mathrm{ref}}$ is recalculated for each geometry. Cross-geometry comparisons are therefore made at matched direct-reference SNR, not at one common absolute instrument noise density. This choice isolates changes in operator geometry and historical content from trivial changes in overall direct-image brightness.

Under an ideal white-noise scaling $\mathrm{SNR} \propto \sqrt{T_{\mathrm{int}}}$, increasing $\mathrm{SNR}_0$ from 100 to 30,000 would correspond to $300^2 = 90{,}000$ times more integration. This is only a scaling dictionary for the dimensionless experiment, not an observability forecast for an existing array. For dimensional orientation, $1M = GM/c^3$. Using the EHT mass scales $4.0 \times 10^6\, M_\odot$ for Sgr A* and $6.5 \times 10^9\, M_\odot$ for M87* [25, 26], $32M$ corresponds to approximately 10.5 minutes and 11.9 days, respectively, while $80M$ corresponds to 26.3 minutes and 29.6 days. These are coordinate-time conversions only; they do not convert $\mathrm{SNR}_0$ into telescope feasibility.

### 3.6 Whitened operators and rank conventions

Let

$$B = C^{-1/2} A \tag{19}$$

We distinguish:

1. exact algebraic rank in symbolic or structurally exact tests;
2. numerical rank under a declared floating-point tolerance;
3. operational rank under a finite-SNR threshold;
4. representation-restricted rank of $A^{(d)} = \mathcal{A}Q_d$ on $\mathcal{S}_d$ (and, separately, secant or tangent stability on a bounded nonlinear class);
5. stable rank $\|B\|_F^2/\|B\|_2^2$.

No paper claim should use the unqualified word "rank" when these notions differ.

# 4 Historical Reach, Dimension, and Recovery

## 4.1 Historical equivalence and resolved information

Two histories are observationally equivalent when their difference lies in the forward null space. For a resolved stack,

$$\ker \mathcal{A}_{g,N} = \bigcap_{n=0}^{N} \ker \mathcal{T}_{g,n} \tag{20}$$

and, for independent whitened order blocks,

$$G_N = A_N^{\mathsf{T}} C_N^{-1} A_N, \qquad G_{N+1} - G_N = A_{N+1}^{\mathsf{T}} C_{N+1}^{-1} A_{N+1} \succeq 0 \tag{21}$$

These identities show that resolving another order cannot remove information already present. They do not guarantee that the new directions are strong enough to matter at finite SNR.

## 4.2 Direct-image old-epoch blindness

Let a finite source representation be separable,

$$q_{\ell k}(r, \phi, t) = \psi_\ell(r, \phi)\tau_k(t) \tag{22}$$

and let

$$\mathcal{W}_n = \{t_o - \Delta_{n,p} : t_o \in \mathcal{O},\ p \in \mathcal{P}_n\} \tag{23}$$

be the discrete retarded-time footprint of order $n$ over the declared observer samples and retained rays.

THEOREM 4.1 — DIRECT-IMAGE OLD-EPOCH BLINDNESS

If $\operatorname{supp} \tau_k \cap \mathcal{W}_0 = \varnothing$, then

$$A_0 q_{\ell k} = 0 \quad \text{for every spatial factor } \psi_\ell \tag{24}$$

For the resolved stack $A_{0:N}$, the same block is identically zero when $\operatorname{supp} \tau_k$ misses $\bigcup_{n=0}^{N} \mathcal{W}_n$. Conversely, if some retained ray of an added order samples a point at which its transfer weight, $\psi_\ell$, and $\tau_k$ are all nonzero, the corresponding stacked column is nonzero.

COROLLARY 4.2 — BLIND-BLOCK DIMENSION

If $m$ compact temporal factors miss the direct footprint and the spatial factor has dimension $d_s$, then

$$\operatorname{null}(A_0 Q) \geq m d_s \tag{25}$$

The result is exact and structural; it is not a statement about a small singular value.

COROLLARY 4.3 — CONTINUUM SUPPORT-NULLITY

Let $\Xi_0$ be the continuum direct-order screen domain and $\mathcal{O}$ an observer-time interval. Define the closed continuum footprint

$$\mathcal{W}_0^{\text{cont}} = \overline{\{t_o - \Delta_{g,0}(\xi) : t_o \in \mathcal{O},\ \xi \in \Xi_0,\ \chi_{g,0}(\xi) w_{g,0}(\xi) \neq 0\}} \tag{26}$$

If $\operatorname{supp} \tau_k \cap \mathcal{W}_0^{\text{cont}} = \varnothing$, then

$$\mathcal{T}_{g,0}(\psi_\ell \tau_k) = 0 \quad \text{almost everywhere on } \Xi_0 \times \mathcal{O} \tag{27}$$

Conversely, a nonzero continuum response requires a positive-measure set of screen–time pairs on which the validity mask, transfer coefficient, spatial factor, and temporal factor are all nonzero; mere set intersection is not sufficient.

The sampled theorem is the finite-operator realization of this continuum support condition. The exactly zero columns measured below are therefore not interpreted as small singular values, while the converse statement remains conditional on nonvanishing spatial and transfer factors. This is the specifically gravitational core of the Mahakal phenomenon studied here: near-critical higher-order delay footprints enter source-time regions that the direct geodesic footprint does not reach. The generic fact that richer inverse models become less identifiable is not new; the geometry-specific content is the lifting of exact epoch-local blind blocks by higher-order null geodesics.

### 4.3 A sufficient spatial bottleneck

In the special separable model $A_n = D_n P$ with one common spatial sampler $P$, the resolved rank obeys

$$\operatorname{rank} A_{0:N} \le \operatorname{rank}(P)\,\operatorname{rank}(D) \tag{28}$$

and any spatial direction in $\ker P$ remains invisible under every delay. This common-sampler proposition is a sufficient bottleneck only. It is not a factorization theorem for the physical Kerr operator, whose orders use different source-plane maps $(r_{n,p}, \phi_{n,p})$.

### 4.4 Finite-SNR reach, dimension, and stable recovery

For the unit-normalized localized probe $q_a$, age information is

$$\mathcal{I}(a) = \|C^{-1/2}\mathcal{A}q_a\|_2^2 \tag{29}$$

The oldest passing center is

$$T_{\text{reach}}(\rho, \text{SNR}_0) = \sup\{a : \text{SNR}_0^2\,\mathcal{I}(a) \ge \rho^2\} \tag{30}$$

Because a threshold mask need not be connected, the primary physical span is the passing run connected to the youngest fully supported probe, $L_{\text{anchor}}$; the full mask and longest passing run are retained as diagnostics.

Historical dimension over a projector $P_T$ is

$$D_{\text{hist}}(T; \rho, \text{SNR}_0) = \#\{i : \text{SNR}_0\,\sigma_i(C^{-1/2}\mathcal{A}P_T Q_d) \ge \rho\} \tag{31}$$

with smooth companion $d_{\text{eff}} = \sum_i \sigma_i^2/(\sigma_i^2 + \lambda)$. Reach and dimension need not move together.

For held-out reconstruction, define

$$E(a) = \frac{\|W_a(\hat{x} - x)\|_2}{\max\{\|W_a x\|_2, \eta\}} \tag{32}$$

The stable anchor-connected span is

$$L_{\text{stable}}^{\text{anchor}}(\epsilon, q) = \sup\left\{T - a_{\text{anchor}} : \Pr_{x,\eta}\left[\sup_{a_{\text{anchor}} \le a \le T} E(a) \le \epsilon\right] \ge q\right\} \tag{33}$$

In the empirical evaluation, probability is over the joint held-out source-history and noise-draw ensemble. Source histories are the independent resampling units; all repeated noise draws associated with one history remain clustered with it.

For a genuinely bounded source set $\mathcal{C}$, the restricted lower secant constant

$$\alpha(A;\mathcal{C}) = \inf_{x_1 \neq x_2 \in \mathcal{C}} \frac{\|C^{-1/2}A(x_1 - x_2)\|}{\|x_1 - x_2\|} \tag{34}$$

provides the standard deterministic stability bound $\|\hat{x} - x_\star\| \leq 2\|C^{-1/2}\eta\|/\alpha$. The theorem does not create identifiability; it quantifies stability once a positive restricted constant exists.

### 4.5 Secondary diagnostics

We separately report level and structure errors, common-subspace errors, throughput versus matched Fisher-sensitivity attenuation, weighted delay quantiles, and the old-age information volume

$$J_{\text{old}}^{(r)} = \int_{a > A_{0,0.999}} \log\left[1 + \mathcal{I}^{(r)}(a)\right] da \tag{35}$$

where $A_{0,0.999}$ is the direct order's 99.9% throughput-weighted age boundary. With dimensionless logarithm and integration over age, $J_{\text{old}}$ has units of $M$. It measures how much finite-SNR sensitivity lies beyond the direct boundary, not whether delay, position, and transfer weight retain their physical pairing.

Finally, attenuation of a known scalar echo train is not intrinsically an instability: for $y(t) = \sum_{n\geq 0} a^n x(t - n\tau)$ with $|a| < 1$, $x(t) = y(t) - ay(t - \tau)$. The difficult obstructions here are finite windows, distributed delays, spatial projection, order mixing, source mismatch, and noise.

## 5 Validated Computational Operator

The physical operator is matrix-free, with explicit forward and adjoint actions in CPU float64. The main-text validation facts are: (i) dense/matrix-free parity and adjoint consistency; (ii) covariance-correct resolved-to-unresolved mixing; (iii) continuum split/merge Gram invariance; (iv) independent AART/kgeo agreement;

(v) an exact Schwarzschild branch; and (vi) one prespecified 12-cell geometry grid. Detailed compatibility incidents, exact/near-null canaries, hashes, and reproduction records are retained in the computational archive rather than repeated here.

AART 2.1.10 supplies the primary Kerr ray maps for $n = 0, 1, 2$ [12]; kgeo supplies an independent cross-tracer. At the reconstruction geometry, source radii agree to $1.082 \times 10^{-12}$, emission times to $2.8 \times 10^{-6} M$, and azimuths after one rigid $\pi/2$ origin rotation to $6.3 \times 10^{-12}$. At $a_\star = 0$, an explicit Schwarzschild circular/plunging branch reproduces $r_+ = 2M$, $r_{\rm ph} = 3M$, and $b_c = 3\sqrt{3}M$ and passes 19 backend checks. No upstream dependency was modified.

The prespecified grid is

$$a_\star \in \{0,\ 0.5,\ 0.9,\ 0.98\}, \qquad i \in \{20^\circ,\ 50^\circ,\ 75^\circ\} \tag{36}$$

Every order and geometry exceeds the declared ray-count threshold. Under the corrected continuum-noise model, the worst split/merge Gram discrepancy is $5.40 \times 10^{-15}$ against $10^{-10}$; all dense/matrix-free, adjoint, mixing-covariance, Gram-monotonicity, weight-semantics, and map-hash checks pass.

## 6 Physical Operator Audit

### 6.1 Reference geometry and distributed delays

The anchor geometry used for detailed reconstruction is

$$a_\star = 0.5, \qquad i = 50^\circ, \qquad n \in \{0, 1, 2\} \tag{37}$$

The valid per-order ray counts are 15597, 8531, and 4179. The delay distributions are broad and overlapping:

$$\begin{aligned} n = 0 &: [0,\ 57.9]\,M, \\ n = 1 &: [46.3,\ 103.4]\,M, \\ n = 2 &: [62.4,\ 120.2]\,M. \end{aligned} \tag{38}$$

At this geometry, AART and kgeo agree on source radius to worst-case $1.082 \times 10^{-12}$ and on emission time to $2.8 \times 10^{-6} M$ after a common time-origin check. The azimuth conventions differ by one rigid $\pi/2$ origin rotation, with order-independent residual $6.3 \times 10^{-12}$.

### 6.2 Prespecified 12-geometry audit and interval semantics

The physical-operator audit reports singular spectra and conditioning on $\mathcal{S}_{224}$, while its historical endpoints use the scalar and 28-direction age-localized source-function probes defined in Section 3.3. It evaluates

$$a_\star \in \{0,\, 0.5,\, 0.9,\, 0.98\}, \qquad i \in \{20^\circ,\, 50^\circ,\, 75^\circ\} \tag{39}$$

for all eight prespecified observation and mechanism arms, using one common $\mathcal{S}_{224}$ construction, eight observer times uniformly spanning $20M$, a quadrature-preserving common count of 1536 rays per order, and an age grid from 0 to $252M$ in $4M$ steps. The 12 cells are deterministic audit points, not a sample from the continuous Kerr parameter space. The float64 physical matrices do not support an algebraic "exact rank" statement; numerical rank is a tolerance decision and operational rank is a finite-SNR decision.

The continuum split/merge verification gives a worst relative Gram discrepancy of $5.40 \times 10^{-15}$ against a $10^{-10}$ criterion. Dense/matrix-free parity, adjoint consistency, mixing-covariance propagation, Gram monotonicity, transfer-weight semantics, ray-map hashes, and grid invariance also pass their prespecified checks.

The age anchor is geometry dependent. At $i = 20^\circ$ and $50^\circ$, the youngest fully supported probe is centered at age zero. At $i = 75^\circ$, the minimum physical delay exceeds the final observer sample, so the youngest fully supported center is 28 – $32M$ depending on spin. Across all arms and SNRs, the longest detectable run differs from the run connected to the anchor in 316 of 1152 rows. The complete threshold mask is therefore part of the result.

### 6.3 Historical extension across the 12-cell grid

At matched direct-reference $\mathrm{SNR}_0 = 100$, the resolved stack extends the anchor-connected detectable span beyond the direct image in all 12 cells. The magnitude is not uniform: the gain is $40M$ at $20^\circ$, $24M$ at $50^\circ$, and one $4M$ grid step at $75^\circ$. The high-inclination cells therefore establish a positive threshold crossing, not a sharply localized additional historical layer. The absolute spans are:

TABLE 3

Direct/resolved anchor-connected spans in $M$ at $\mathrm{SNR}_0 = 100$. The repeated gain values in Fig. 3a are quantized on the $4M$ age grid; the absolute endpoints vary with spin at high inclination.

| | 20° | | 50° | | 75° | |
|---|---|---|---|---|---|---|
| $a_\star$ | Direct | Resolved | Direct | Resolved | Direct | Resolved |
| 0.00 | 20 | 60 | 60 | 84 | 108 | 112 |
| 0.50 | 20 | 60 | 60 | 84 | 112 | 116 |
| 0.90 | 20 | 60 | 60 | 84 | 112 | 116 |
| 0.98 | 20 | 60 | 60 | 84 | 112 | 116 |

The finite-SNR spectral improvement is also substantial. Across the 12 cells, the median operational rank increases from 153 for the direct image to 202 for the resolved stack, a 32% increase. The median positive-support condition number decreases from $9.00 \times 10^8$ to $7.84 \times 10^5$, an improvement by approximately $1.15 \times 10^3$. The median old-age innovation in Eq. (35) increases from $8.29M$ to $40.72M$, a factor of 4.91. Here $A_{0,0.999}$ is the direct order's 99.9% throughput-weighted age boundary and $J_{\mathrm{old}}$ has units of $M$. Absolute reach grows with inclination because the direct image itself samples a broader retarded-time range. The incremental value of higher orders therefore falls from $40M$ to $24M$ to $4M$: at high inclination the direct channel already occupies much of the available delay span.

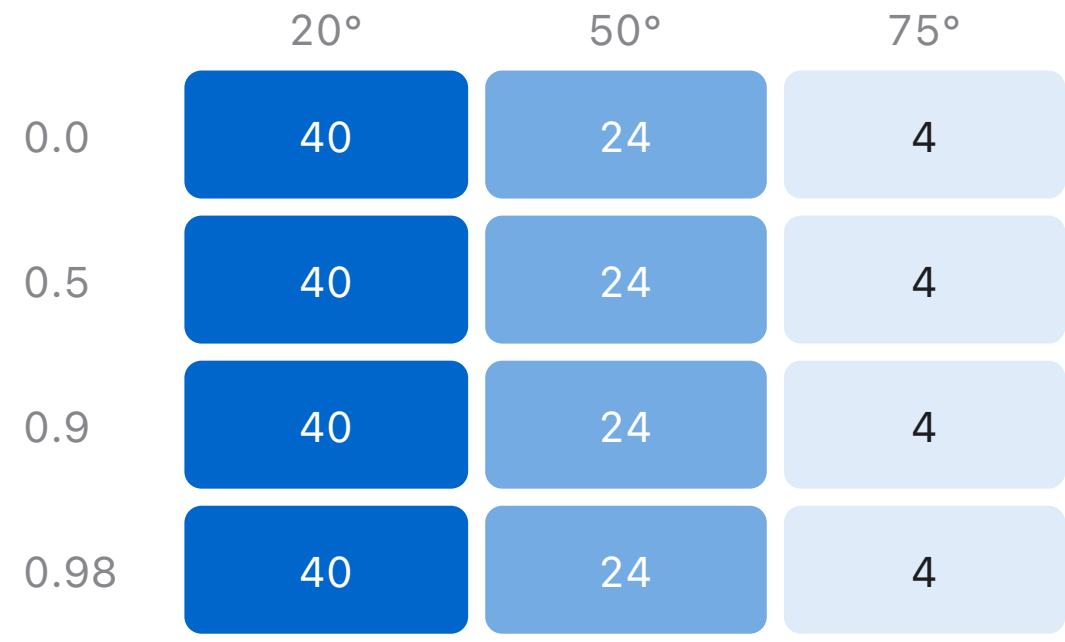

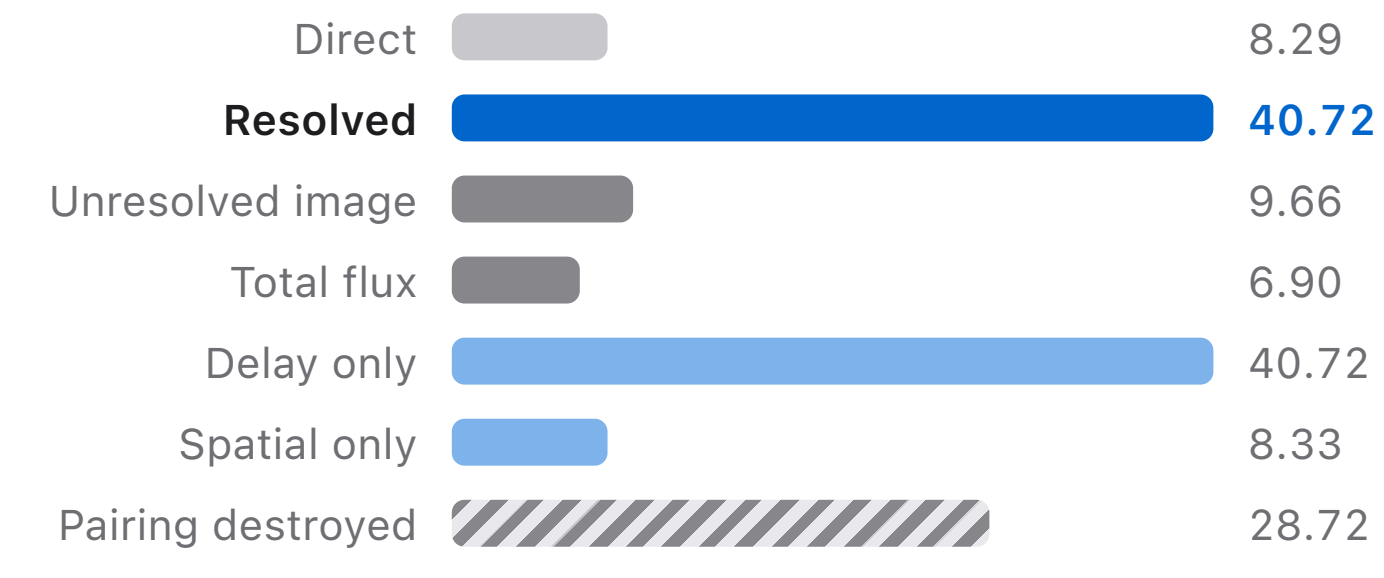

FIGURE 3

Information added across the prespecified 12-cell Kerr/Schwarzschild grid at $\text{SNR}_0 = 100$. Each cell is deterministic rather than a draw from a population. The pairing-destroyed control retains old-delay marginals but destroys their association with source position and transfer weight. It therefore tests physical fidelity, not merely conditioning.

## 6.4 Mechanism decomposition

The 12-cell mechanism result sharpens the original reference-geometry interpretation. At $\text{SNR}_0 = 100$, the delay-only arm has the same median oldest detectable probe and the same median $J_{\text{old}}$ as the full resolved stack: $84M$ and $40.72M$. The spatial-only arm has median reach $60M$ and median $J_{\text{old}} = 8.33M$, closely tracking the direct image. Thus, on the declared class and after the direct spatial map has already supplied substantial injectivity, the added historical endpoint is carried primarily by the physical retarded-time field.

The unresolved image is not equivalent to the resolved stack over the full grid. Its median oldest detectable probe is $60M$, and its median $J_{\text{old}} = 9.66M$, only modestly above the direct value and just 23.7% of the resolved value $40.72M$. At the reference geometry, unresolved reach happens to remain close to resolved reach; the 12-cell result shows that this is not a safe generalization. Total-flux collapse is more destructive still, with median operational rank 13 and median $J_{\text{old}} = 6.90M$. Explicit order labels are therefore nearly dispensable for the oldest scalar threshold at high inclination but load-bearing for the amount and morphology of recoverable historical information.

The pairing-destroyed control has median $J_{\mathrm{old}} = 28.72M$. It retains 70.5% of the resolved information volume, or 63.0% of the resolved increment above direct imaging. This does not refute $J_{\mathrm{old}}$; it fixes its meaning. The statistic measures old-age sensitivity volume, while a permutation can preserve old-delay marginals and attach them to the wrong spatial directions. Physical interpretation requires the correctly paired Kerr operator and held-out recovery, not $J_{\mathrm{old}}$ or conditioning alone.

### 6.5 Throughput, matched sensitivity, and tail convergence

At the reference geometry, the continuum-corrected matched-window analysis gives

$$\begin{array}{lcc} & 0 \to 1 & 1 \to 2 \\ \Gamma_{\mathrm{throughput}} & 4.272 & 3.784 \\ \Gamma^{\mathrm{matched}}_{\mathrm{sensitivity}} & 2.486 & 2.120 \end{array} \tag{40}$$

across all 19 prespecified matched fractions. Brightness therefore decreases more steeply than source-class-specific matched sensitivity, but the difference is finite and observable dependent; it is not a universal conversion law.

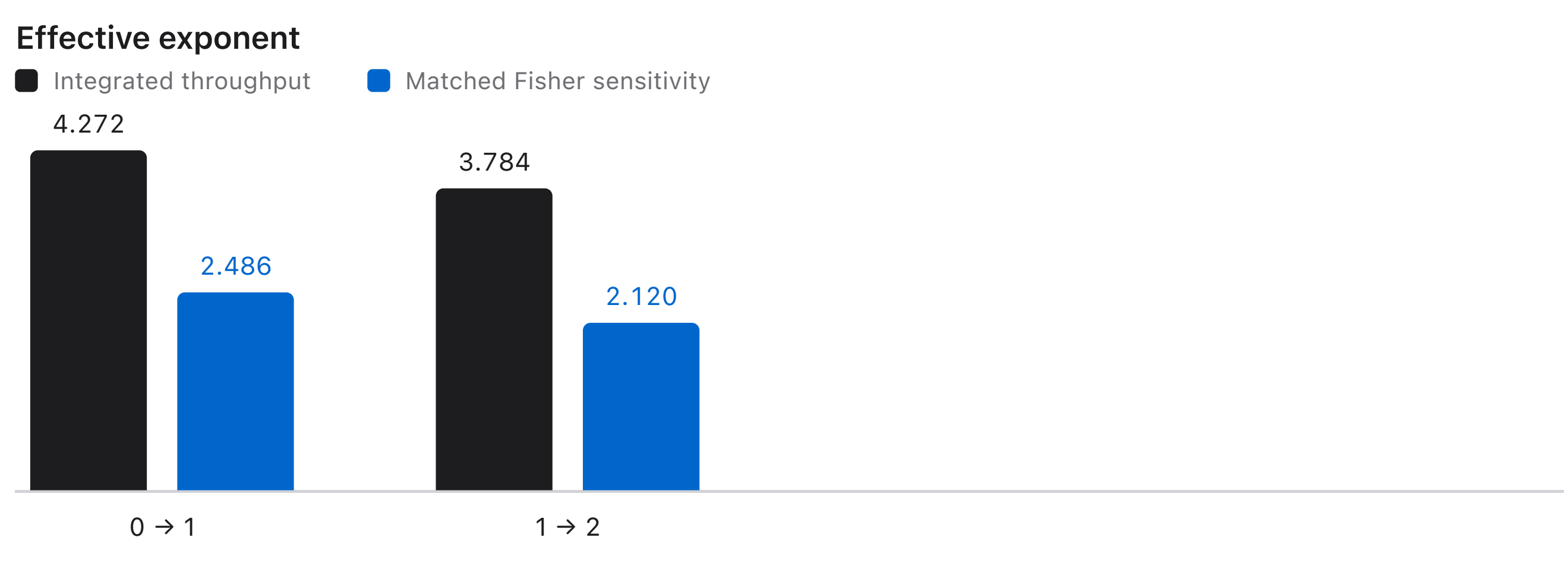

FIGURE 4
Reference-geometry effective exponents: integrated throughput attenuates more steeply than matched age-localized Fisher sensitivity.

The sampled maximum retarded age remains an unstable tail statistic under refinement and is not used as historical depth. Weighted delay quantiles and the quadrature-consistent Gram operator converge, while the raw maximum is retained only as a retired diagnostic.

## 7 Compact Support Verifies Exact Old-Epoch Blindness

Global DCT modes span the whole source-time window; fitting their coefficients cannot ask whether one localized epoch is directly observed or extrapolated from another. We therefore replace them by compactly supported degree-one B-splines while retaining the same rays and mirrored dimensions. Theorem 4.1 then predicts exact zero column blocks whenever a temporal support misses the direct footprint.

For $L_{224}$, three missing temporal functions crossed with 28 spatial functions give exactly $3 \times 28 = 84$ zero columns. For $L_{448}$, $7 \times 28 = 196$. In the richest class, $523 + 517 = 1040 < 1056$, leaving 16 columns that are nonzero but linearly dependent; exact zero support and linear dependence are different mechanisms.

Restricting to old compact temporal functions and removing the spatially constant level gives the old-structure subspace:

| Class | Direct | Resolved | Order-summed image |
|---|---:|---:|---:|
| $L_{224}$ | 0 | 38 | 21 |
| $L_{448}$ | 0 | 93 | 46 |
| $L_{1056}$ | 0 | 185 | 91 |

(41)

The direct largest singular value on this subspace is $1.8 \times 10^{-14}$, numerically zero. Higher-order delay footprints therefore lift tens to hundreds of directions in a subspace the direct image does not sample at all; roughly half the operational directions survive complete order summation. No truth is drawn and no estimator is fitted in this section.

TABLE 4

Direct-image compact-support audit at $a_\star = 0.5$, $i = 50^\circ$. "Mirrored global rank" is the direct-image rank on the equal-dimensional global-DCT representation used as the compact-support comparison.

| Class | Dimension | Direct rank | Exactly zero direct columns | Mirrored global rank |
|---|---|---|---|---|
| $L_{224}$ | 224 | 140 | 84 | 224 |
| $L_{448}$ | 448 | 252 | 196 | 411 |
| $L_{1056}$ | 1056 | 523 | 517 | 911 |

## 8 Nested Source-Class Stress

The physical audit on $\mathcal{S}_{224}$ establishes a finite-representation result, not injectivity of the underlying continuum operator. To expose that boundary directly, the source model is enriched on three prespecified anchors,

$$(a_\star, i) \in \{(0, 20^\circ),\ (0.5, 50^\circ),\ (0.98, 75^\circ)\} \tag{42}$$

using the nested function spaces

$$\mathcal{S}_{224} \subset \mathcal{S}_{448,T}, \qquad \mathcal{S}_{224} \subset \mathcal{S}_{528,S}, \qquad \mathcal{S}_{448,T} \subset \mathcal{S}_{1056,ST} \tag{43}$$

Their radial, real-azimuthal, and temporal dimensions are $(4, 7, 8)$, $(4, 7, 16)$, $(6, 11, 8)$, and $(6, 11, 16)$, respectively. Temporal and azimuthal enrichment retain the parent modes as literal prefixes. Refining the cubic radial B-splines moves individual knots, so nesting is verified at the level of function-space containment; the worst parent-to-child projection residual is $2.73 \times 10^{-14}$. Dense/matrix-free parity, the adjoint, Gram monotonicity, class nesting, and nondecreasing numerical rank all pass their prespecified checks.

Temporal enrichment is the stronger identifiability stress. On $\mathcal{S}_{448,T}$ the direct operator is rank deficient at all three anchors—$348/448$, $411/448$, and $444/448$—whereas the resolved stack remains full rank at the first two and is $447/448$ at high inclination. Spatial enrichment to $\mathcal{S}_{528,S}$ is gentler: the direct operator is deficient only at the low-inclination Schwarzschild anchor $(522/528)$, while the resolved stack is full rank at all three anchors. On $\mathcal{S}_{1056,ST}$, the direct operator loses 109–265 dimensions and the resolved stack loses 11–35.

The historical endpoint barely moves. Across the physical and mechanism arms, class enrichment changes the localized reach by no more than one $4M$ grid step, while the smallest positive singular value falls by roughly five orders of magnitude over the ladder and the representative positive-support condition number rises from $7.04 \times 10^5$ to $1.04 \times 10^{11}$. This is the clearest computational demonstration that historical depth, historical dimension, and stability are different quantities: a channel can continue to touch old epochs while determining a shrinking fraction of a richer source model.

The enriched-space audit does not claim stable reconstruction on $\mathcal{S}_{448,T}$, $\mathcal{S}_{528,S}$, or $\mathcal{S}_{1056,ST}$. It establishes why every reconstruction statement below must name its representation and why historical reach cannot be used as a proxy for recoverable historical dimension.

TABLE 5

Source-class stress at the reconstruction geometry $a_\star = 0.5$, $i = 50°$. Rank is numerical rank of the whitened coefficient operator. "Localized directional reach" is the deepest age at which at least one normalized localized direction clears threshold. It differs from the anchor-connected scalar-probe span in Table 3, which is why the same geometry reads $60/84M$ there and $60/92M$ here. Neither is a held-out reconstruction span.

| Space | Dimension | Direct rank | Resolved rank | Direct reach | Resolved reach |
|---|---|---|---|---|---|
| $\mathcal{S}_{224}$ | 224 | 224 | 224 | $60M$ | $92M$ |
| $\mathcal{S}_{448,T}$ | 448 | 411 | 448 | $60M$ | $92M$ |
| $\mathcal{S}_{528,S}$ | 528 | 528 | 528 | $60M$ | $92M$ |
| $\mathcal{S}_{1056,ST}$ | 1056 | 911 | 1045 | $60M$ | $92M$ |

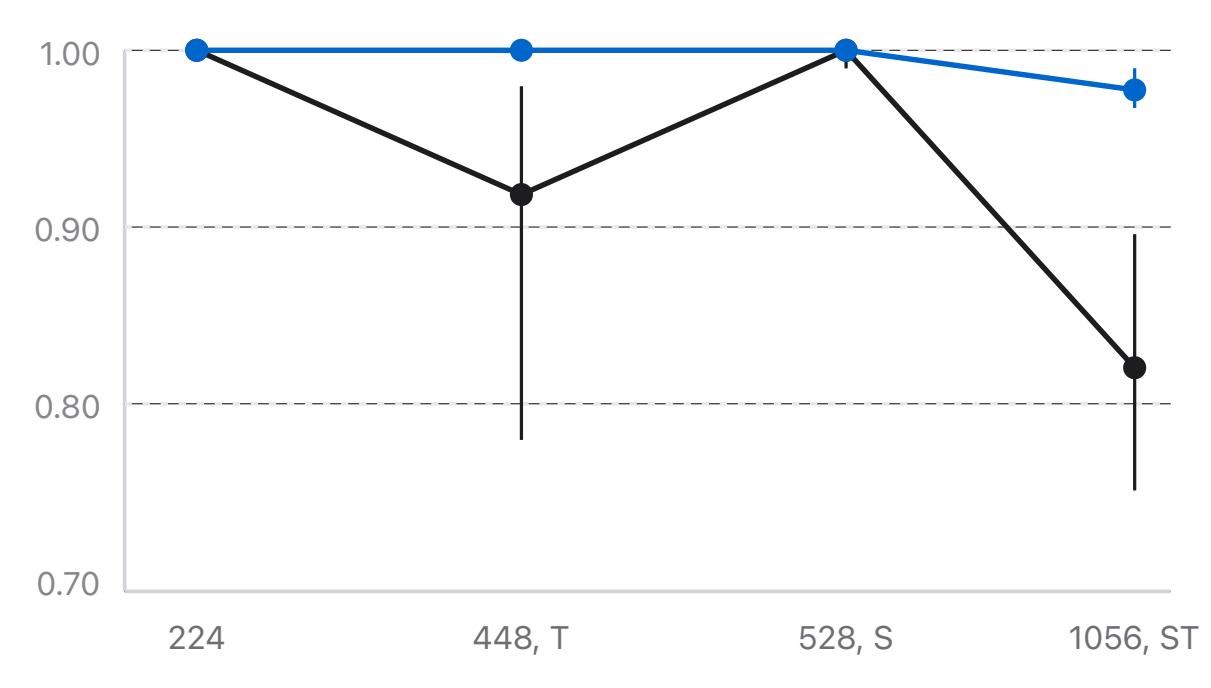

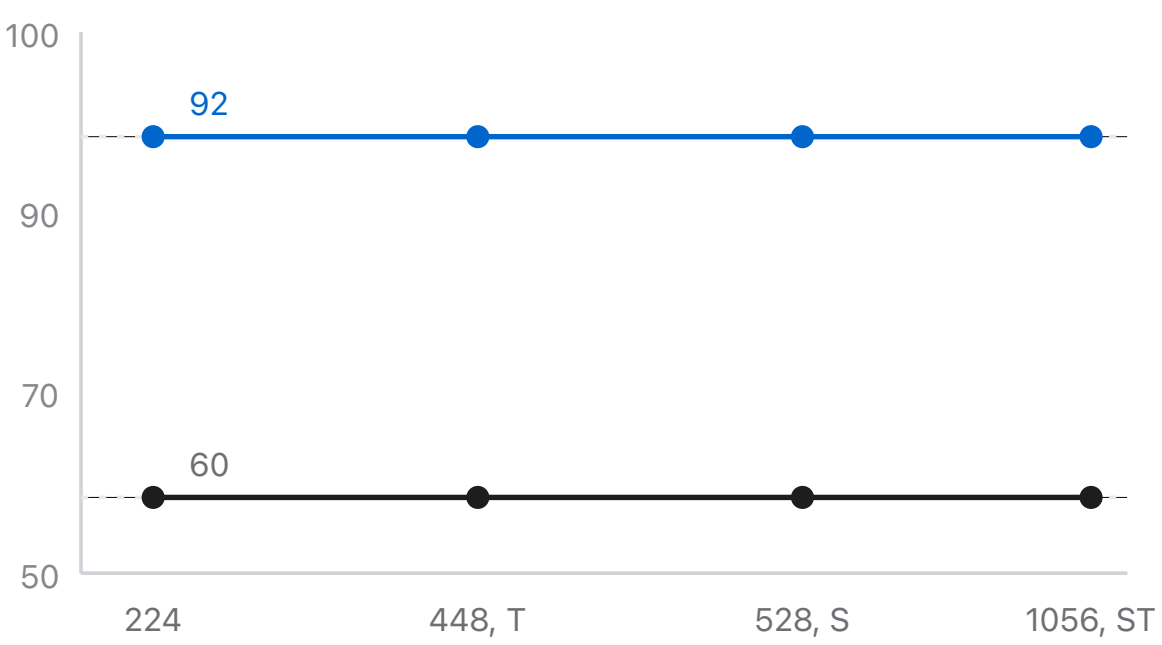

Direct $n = 0$ Resolved $n = 0, 1, 2$

FIGURE 5

Source richness separates historical reach from recoverable dimension. Left: median numerical-rank fraction over the three anchor geometries, with min–max bars. Right: at the reconstruction anchor, localized reach remains $60M$ direct and $92M$ resolved across the class ladder even as null directions appear. The long archive therefore persists while the fraction and conditioning of recoverable directions degrade.

## 9 Held-Out Inverse Result I: Age-Local Emissivity Level

The first held-out inverse experiment asks whether the additional finite-SNR information produces a longer stable history on the original 224-dimensional representation. The geometry is fixed at $a_\star = 0.5$, $i = 50^\circ$. The primary estimators are truncated SVD and ridge/Tikhonov. They are non-Bayesian spectral estimators with no learned or stochastic source prior, but the fixed representation, known geometry, and validation-selected regularization remain structural assumptions.

The sealed bank contains 640 histories spanning in-representation and off-grid regimes, including a held-out flare family. Every record was committed by content hash before scoring; the main run regenerated the bank, verified split disjointness, and reused validation-selected hyperparameters. The primary endpoint is the anchor-connected stable span at $(\epsilon, q) = (0.25, 0.95)$ and $\mathrm{SNR}_0 = 100$. Empirically, $q$ is evaluated over the joint held-out truth–noise ensemble, with truths as the independent bootstrap clusters.

TSVD and ridge both reproduce the $32M$ gain, four times the prespecified $8M$ materiality threshold. A paired bootstrap over truths gives old-band normalized error reduction 0.332 with lower bound 0.330 and absolute reduction 0.429 with lower bound 0.419. A like-for-like comparison on the direct channel's own data-supported subspace also improves: $0.986 \to 0.579$ for TSVD and $0.850 \to 0.445$ for ridge.

At this geometry the $n = 2$ delay footprint adds roughly $62M$ beyond the direct-order edge, while the stable level span grows by $32M$; approximately 52% of the reachable extension is converted into the registered stable span in this one experiment.

### 9.1 Why this is not morphology reconstruction

The prespecified age-window norm is dominated by the spatially constant component: 98.4% of the held-out norm lies in the level subspace. Decomposing $x = P_{\text{level}}x + P_{\text{structure}}x$ and normalizing each component by its own truth norm gives

| | Direct | Resolved |
|---|---|---|
| all-age level error | 0.266 | 0.028 |
| all-age structure error | 0.927 | 0.466 |
| old-band level error | 0.653 | 0.217 |
| old-band structure error | 1.169 | 1.141 |

(44)

The structure-only stable span is therefore $0M$ for both arms at $\text{SNR}_0 = 100$. Nonzero structure span first appears at $\text{SNR}_0 = 30{,}000$, where the direct and resolved spans are $40M$ and $76M$; the onset SNR is not lowered. Under ideal white-noise scaling this operating point is 300 times larger in normalized amplitude, or formally $9 \times 10^4$ times longer in integration than the reference. It is a secondary high-SNR structural result, not an observational forecast.

TABLE 6

Stable baseline-inclusive age-local emissivity span in the sealed bank. Values are in $M$.

| Regime | Direct | Resolved | Difference |
|---|---|---|---|
| In-class, fitting families | 48 | 80 | +32 |
| In-class, held-out flare family | 48 | 80 | +32 |
| Mild off-grid held-out family | 48 | 80 | +32 |
| Severe off-grid fitting families | 0 | 0 | 0 |

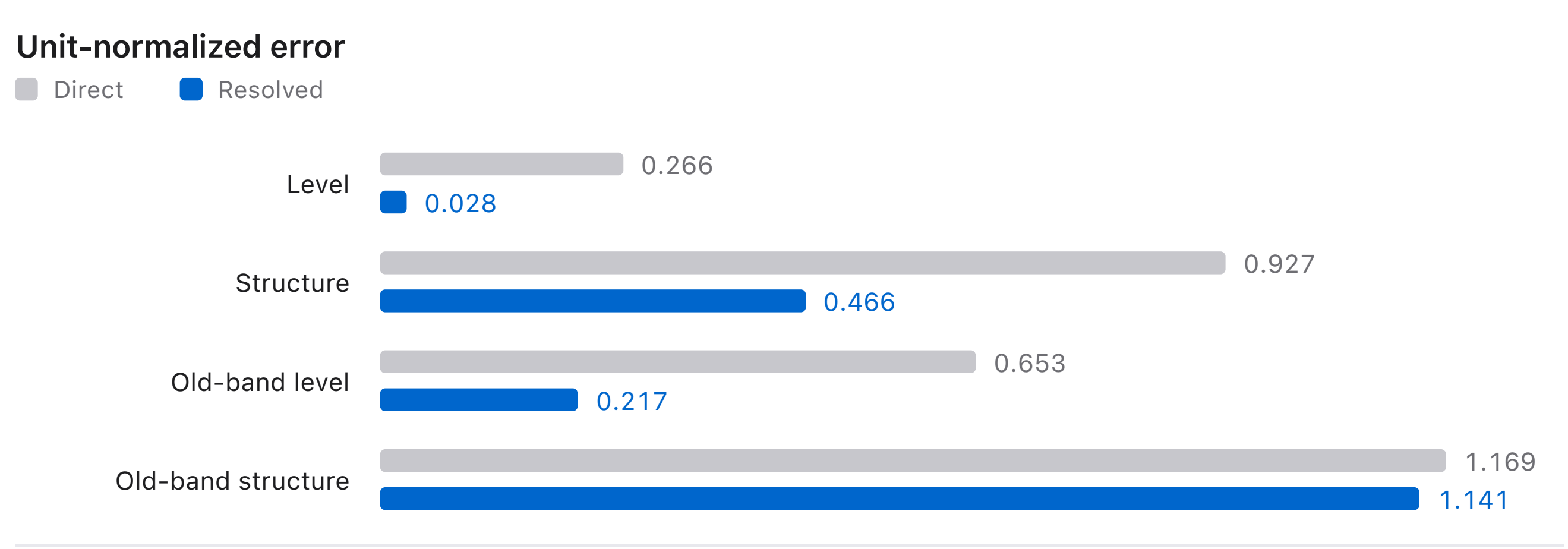

Bars scaled to 1.30; unit error is the point where a reconstruction carries no usable information.

FIGURE 6

Level and structure are different historical objects. At the reference SNR, resolved orders strongly improve the spatially constant level and all-age structure, but old-band structure remains above unit-normalized error for both arms.

Severe off-grid histories remain a genuine negative result: the exact projection reaches the tolerance, but neither direct nor resolved reconstruction does. Posterior uncertainty is withdrawn because the joint calibration criterion failed; probabilistic estimators are retained only as point estimates. This experiment therefore supports stable emissivity-level history on a bounded representation, not robust or calibrated movie recovery.

## 10 Held-Out Inverse Result II: Resolution-Aware Historical Morphology

A pixel-wise movie metric is ill defined when the source model, evaluation grid, and physical feature widths do not resolve the same objects. The morphology program therefore began with a source-only audit, before any ray map or observation operator was imported. Across 169 sources and three refinement grids, the two-hotspot family contained resolved pairs, stable blends, and topology-ambiguous states. On states stably classified as multi-resolved, projection into the original contrast class merged 29.9% of pairs; radial enrichment to a 896-dimensional contrast class reduced that rate to 13.6%. This audit fixed the historical object before reconstruction rather than after a failure.

Each truth-age state is assigned one of five labels: single resolved, multiple resolved, blended, dead, or ambiguous between the two finest grids. The label selects a normalized error: optimal set assignment for resolved states, centroid/scale/mode descriptors for blends and ambiguous states, and amplitude for dead states. No state is excluded. Two targets are always reported: **class-conditional**, which compares to the best in-class source object, and **physical end-to-end**, which compares to the analytic source.

TABLE 7

Primary morphology design and inference at $\mathrm{SNR}_0 = 100$.

| Target | Estimator | Histories | Draws | Median reduction | 95% low |
|---|---|---|---|---|---|
| Physical end-to-end | TSVD | 60 | 4 | 0.133 | 0.101 |
| Physical end-to-end | Ridge | 60 | 4 | **0.164** | 0.116 |
| Class-conditional | TSVD | 60 | 4 | 0.160 | 0.092 |
| Class-conditional | Ridge | 60 | 4 | 0.158 | 0.114 |
| Non-dead physical | TSVD | 60 | 4 | 0.142 | > 0.05 |
| Non-dead physical | Ridge | 60 | 4 | 0.158 | > 0.05 |

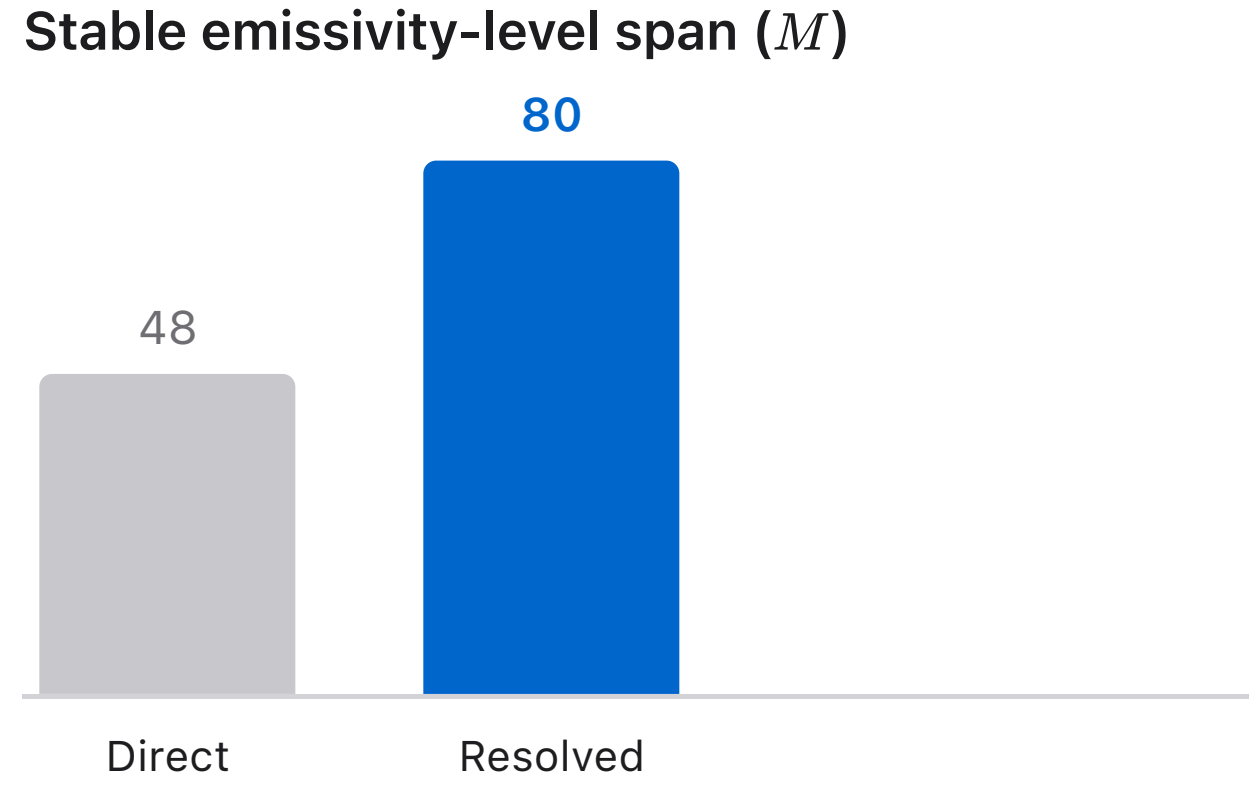

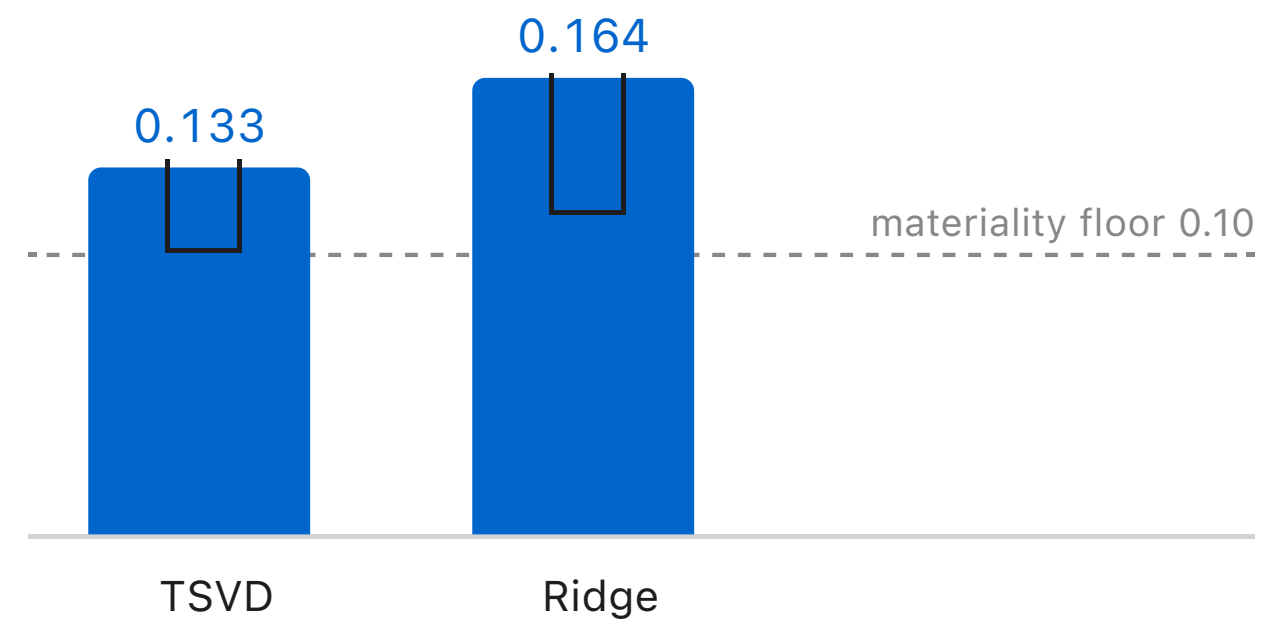

FIGURE 7
The two held-out historical inverse results measure different objects and units. Left: stable baseline-inclusive emissivity-level span. Right: paired morphology-error reduction against the prespecified 0.10 materiality floor.

### 10.1 Sealed reduced-scope main

The held-out main uses one geometry, the primary 896-dimensional radial-enriched contrast class, 60 fresh source histories, and four paired noise draws. The count is smaller than an earlier 96-history/eight-draw authorization, but the reduced scope was committed before any truth was drawn. Hyperparameters were inherited unchanged from validation; the endpoint runner contains no sweep. The primary comparison is resolved versus direct at $\mathrm{SNR}_0 = 100$.

The contemporaneous record does not state why the scope was reduced, so no retrospective resource rationale is supplied. The independent resampling unit is the source history; all four noise draws travel with the same history. Materiality was fixed before the bank at a median paired reduction of at least 0.10 and a 95% lower bound of at least 0.05.

All primary physical and class-conditional point estimates clear the declared 0.10 materiality floor, and their lower bounds clear 0.05. The non-dead companion also remains material. The order-summed image reaches at most 0.015 and total flux is non-material or negative, so the gain is attributable to resolving the orders rather than merely collecting their photons. In the representation-limited 448-dimensional control class, only ridge reaches materiality.

### 10.2 Four qualifications that are part of the result

**Family heterogeneity.** Five of 12 family-estimator cells are material on the physical target and four of 12 are material on both targets. The largest effects occur for flare birth-motion-decay and the $m = 2$ structural mode; the simplest circular-hotspot family is negative under both estimators. The aggregate is therefore a real average over a heterogeneous set, not a uniform source-family law.

A post-hoc leave-one-family-out sensitivity check recomputes the equal-family mean of the six family medians without retuning. The full means are 0.169 (TSVD) and 0.196 (ridge); across the six omissions they range from 0.127 to 0.213 and from 0.159 to 0.239. Every omission leaves a positive aggregate, including omission of the flare or $m = 2$ family. This post-hoc analysis does not alter the preregistered claim; it shows only that the aggregate sign is not created by one family alone.

---

**Per-family physical end-to-end reduction: what the aggregate is made of**

| 5 / 12 | 4 / 12 | 0.169 · 0.196 |
| --- | --- | --- |
| family-estimator cells material on the physical target | material on both targets | equal-family mean, TSVD and ridge |

- Largest effects: flare birth-motion-decay, $m = 2$ structural mode
- Negative under both estimators: the simplest circular-hotspot family
- Leave-one-family-out means: 0.127–0.213 (TSVD), 0.159–0.239 (ridge)

FIGURE 8

Per-family physical end-to-end morphology reduction. Ten truths per family make the family-level intervals wide, so this figure qualifies the aggregate rather than supporting strong per-family claims.

**Multi-feature recovery is negative.** No stable multi-resolved family-estimator cell reaches materiality. The resolved assignment cost is approximately 1.21–1.30, where 1.0 corresponds to one whole feature being unmatched. The method often recognizes structured morphology and improves one retained component, but it does not reliably recover the resolved pair.

**Stable morphology interval is negative.** For every arm, estimator, class, and tested SNR,

$$L_{\text{stable,morph}}(\epsilon = 0.25,\ q = 0.95) = 0M \tag{45}$$

At the reference SNR, the resolved arm's mean within-tolerance reach is slightly lower than the direct arm's, $3.06M$ versus $3.32M$; at tenfold SNR the ordering reverses, $3.45M$ versus $2.92M$. Neither case yields a nonzero 95%-reliable interval. The result is a reduction in aggregate historical morphology error, not a temporally coherent recovered movie.

**The direct baseline is partly saturated.** Approximately 55–57% of direct-image states sit at the morphology metric's ceiling. The resolved arm reduces the frequency of complete failures, but its absolute all-state error remains about 0.60–0.65 on a scale whose worst case is one. This is meaningful rescue from catastrophic error, not generally accurate recovery.

## 11 What the Evidence Supports

The computational program now supports a hierarchy of claims rather than one slogan.

TABLE 8

Scientific dispositions of the main claims.

| Claim | Disposition | Evidence |
|---|---|---|
| Higher orders extend finite-SNR historical observability | Supported | Positive in all 12 prespecified geometries, with gains $40M$, $24M$, and $4M$ by inclination; rank and $J_{\rm old}$ increase. |
| The effect is purely spatial remapping | Rejected | Flattening delays removes most historical gain; spatial-only remains near direct. |
| Long reach implies rich identifiability | Rejected | Source enrichment leaves reach nearly fixed while rank fraction and $\sigma^+_{\rm min}$ collapse. |
| Stable emissivity-level history is extended | Supported, bounded | $48M \to 80M$ on the sealed $\mathcal{S}_{224}$ bank. |
| Moderate-SNR old morphology is stably recovered | Rejected | Structure-only stable span and morphology stable interval are both zero. |
| Aggregate morphology inference improves | Supported, bounded | 13.3–16.4% sealed physical end-to-end reduction; complete order summation is non-material. |
| Two-feature historical tracks are recovered | Rejected | Assignment cost remains above one unmatched feature and no multi-feature cell is material. |
| A historical movie of real data is recovered | Not tested | No real data, geometry mismatch, interferometric projection, or order-leakage study. |

Five methodological lessons are transferable beyond this specific operator. First, the oldest ray is not recoverable depth; weighted temporal support and finite-SNR sensitivity are. Second, rank is a property of an operator and a source representation, not of the spacetime alone. Third, better conditioning can be produced by a nonphysical permutation, so conditioning is not evidence of physical content. Fourth, average error reduction and stable historical interval are different endpoints. Fifth, the source object must be defined at a declared spatial, temporal, and representation resolution before a reconstruction is judged.

## 12 Limitations

1 **Finite source spaces and synthetic families.** Every rank, nullity, and recovery result is conditional on a declared representation or held-out bank; no continuum injectivity or GRMHD-history claim is made.

2 **Three orders and ideal order separation.** The operator retains $n = 0, 1, 2$. Complete order summation retains only 23.7% of the resolved old-age innovation and yields no material morphology gain; no continuous leakage sweep is included.

3 **Ideal image-domain instrument.** The study assumes monochromatic scalar intensity, known geometry, and direct image access. Sparse Fourier sampling, closure quantities, calibration, scattering, polarization, and spectral channels are absent.

4 **One reconstruction geometry.** Both held-out programs use $a_\star = 0.5$, $i = 50^\circ$. Multi-geometry observability is not multi-geometry recovery.

5 **Normalized SNR.** $\mathrm{SNR}_0$ is tied to a direct-reference operator response, not a telescope image SNR or observing forecast.

6 **Finite temporal resolution.** Historical spans use $h = 3M$ probes and a $4M$ age grid. No independent probe-width sweep is performed; the $4M$ high-inclination gain is one grid quantum.

7 **Level-dominated stable-span result.** The $48M \to 80M$ endpoint is baseline inclusive and 98.4% level by norm; old-band structure remains unrecovered at the reference SNR.

8 **Morphology is aggregate and heterogeneous.** The 13.3–16.4% reduction is not reliable two-feature recovery, has zero stable morphology interval, and partly reflects fewer ceiling-saturated failures under the resolved arm.

9 **Reduced-scope morphology main.** The authorized 96-history/eight-draw design was reduced to 60/four before truth generation. The contemporaneous record gives no rationale; the consequence is wider family-level uncertainty.

10 **No calibrated posterior uncertainty.** The preregistered joint calibration criterion failed, so probabilistic estimators are retained only as point estimators.

## 13 Discussion

The results separate three questions that a single depth number cannot answer. Higher-order channels increase historical **reach**: they place measurable sensitivity at source ages beyond the direct footprint. Compact support then shows why that reach is gravitationally distinctive: higher-order delay footprints lift exact epoch-local null blocks of the direct image. Source enrichment simultaneously reduces historical **dimension**: the archive stays long while rank fraction and $\sigma^{+}_{\min}$ collapse. Finally, held-out experiments test historical **recovery**: some bounded objects improve, but no stable morphology movie emerges.

Retarded-time diversity carries the historical endpoint, while order-dependent spatial remapping changes the information geometry. The order-summed control makes the practical condition explicit. Its median $J_{\text{old}}$ is $9.66M$ versus $40.72M$ for resolved orders, and its held-out morphology effect is non-material. The present positive result is therefore an ideal order-resolution result. Photon-ring overlap and any realistic classification leakage must be treated in a separate instrument model rather than inferred from the two endpoints studied here.

The pairing-destroyed control sharpens another boundary. It retains 70.5% of the resolved $J_{\text{old}}$ despite destroying the physical association among delay, source position, and weight. Thus $J_{\text{old}}$ measures old-age information volume, not physical fidelity. Fidelity enters only through the paired Kerr operator and held-out targets. Likewise, a decrease in aggregate morphology error is not a stable recovered history: the stable interval is zero and two-feature assignment remains worse than one unmatched feature.

This is the substantive meaning of the Mahakal phenomenon in this paper. Near-critical geometry creates source-time access that the direct image lacks, but richer historical questions expose the null and weak directions hidden by global low-dimensional models. The result is neither “the past is stored” nor “the past is absent.” It is a measured, source-class-dependent information boundary.

## 14 Conclusion

Near-critical black-hole null geodesics form a distributed retarded-time archive. Across a prespecified 12-cell Kerr/Schwarzschild grid, separately resolving the direct and first two higher-order bands extends the anchor-connected historical span, with gains of $40M$, $24M$, and one $4M$ grid step as inclination increases. Compact temporal support proves that some old-source blocks are exactly absent from the direct image and shows that higher-order footprints lift part of them.

The archive is not equivalent to a movie. Enriching the source model leaves reach nearly fixed while historical dimension and conditioning collapse. On one sealed bank, higher orders extend stable baseline-inclusive emissivity-level span from $48M$ to $80M$, but old-band structure remains unrecovered at the reference SNR. On a second sealed bank, ideal order resolution reduces aggregate physical morphology error by 13.3–16.4% under two classical estimators. Complete order summation does not reproduce the gain; the effect is heterogeneous, does not recover resolved feature pairs, and yields no stable morphology interval. These recovery results are therefore best-case, known-geometry image-domain benchmarks rather than observational forecasts.

> The Mahakal phenomenon is a bounded creation–destruction relation: higher-order geodesics create access to earlier source structure, while localization and source enrichment expose the dimensions that cannot be stably recovered.

# A Proof Details

## A.1 Resolved intersection and Gram monotonicity

For the stacked operator, $A_{0:N}h = 0$ if and only if every block $A_n h = 0$, which proves Eq. (20). Under block-diagonal independent covariance,

$$G_{N+1} = G_N + A_{N+1}^{\mathsf{T}} C_{N+1}^{-1} A_{N+1} \tag{46}$$

and the added term is positive semidefinite.

## A.2 Direct-image old-epoch blindness

A row indexed by $(t_o, p)$ acts on $q_{\ell k}$ as

$$(A_n q_{\ell k})_{t_o,p} = w_{n,p}\, \psi_\ell(r_{n,p}, \phi_{n,p})\, \tau_k(t_o - \Delta_{n,p}) \tag{47}$$

If $\operatorname{supp} \tau_k \cap \mathcal{W}_n = \varnothing$, the temporal factor is zero in every row, so the complete column vanishes. The resolved statement follows blockwise. If one retained row has all three factors nonzero, that row proves the stacked column is nonzero. Crossing $m$ missing temporal factors with $d_s$ spatial factors gives $md_s$ zero columns and hence nullity at least $md_s$.

For Theorem 4.3, separation from the closed footprint implies $\tau_k(t_o - \Delta_{g,0}(\xi)) = 0$ for every supported screen–time pair, so the continuum response vanishes almost everywhere. Conversely, a nonzero $L^2$ response requires a positive-measure set on which the full product in Eq. (2) is nonzero; an isolated intersection of supports does not suffice.

## A.3 Restricted stability

Let $e = \hat{x} - x_\star$. The residual condition and triangle inequality give $\|C^{-1/2}Ae\| \leq 2\|C^{-1/2}\eta\|$. Applying the lower secant constant on the bounded class gives $\alpha(A;\mathcal{C})\|e\| \leq \|C^{-1/2}Ae\|$ and the stated bound.

## A.4 Information-tail implication

If $\|C_n^{-1/2}A_n\| \leq C_0 e^{-\Gamma n}$, then

$$\|G_\infty - G_N\| \leq \sum_{n>N} C_0^2 e^{-2\Gamma n} = \frac{C_0^2 e^{-2\Gamma(N+1)}}{1 - e^{-2\Gamma}} \tag{48}$$

This is a sufficient bound; the observed matched-sensitivity exponent remains source- and measurement-dependent.

## B Reproducibility and Evidence Lineage

All physical operators and scientific linear algebra are evaluated in CPU float64 under a pinned single-threaded numerical environment. Geometry maps, source spaces, age grids, arms, rank conventions, SNR grids, thresholds, splits, and estimator grids were fixed before primary evaluations. Every held-out source carries a content hash, and every main run verifies split disjointness and reuses validation-selected hyperparameters.

The release is tied to a canonical artifact freeze and claim ledger that maps numerical statements to table rows and columns. Mechanical failures remain visible after adjudication rather than being rewritten as passes. Detailed gate, amendment, retired-bank, and platform-reproduction records are provided in the computational archive.

The morphology main is a reduced-scope sealed experiment: 60 histories and four paired draws were fixed before the bank was generated, rather than the authorized 96/eight design. The contemporaneous record does not state the reason for the reduction. Its scoring stage was clean, reproduced all committed histories, and used no hyperparameter sweep. The reduction is treated only as a scope deviation that widens uncertainty.

### AI-Assisted Manuscript Preparation

OpenAI ChatGPT was used under author direction to assist with manuscript restructuring, language editing, literature organization, and document-quality checks. The authors independently reviewed and verified the equations, citations, numerical claims, and final text. No AI-generated illustration is included as a scientific figure in this version.

### C Data and Code Availability

No proprietary observational data are used. The versioned computational archive associated with this manuscript contains source code, fixed configurations, ray-map digests, canonical numerical tables, deterministic figure builders, the claim ledger, and content-hash manifests required to regenerate the reported results. Every numbered figure in this manuscript is either generated deterministically from canonical numerical values or is a mathematical schematic constructed from those values.